# Estimulando o ensino de ciências através da compostagem

**Michele Cristina Muniz[1], Emanuele Vitoria da Silva[2], João Pedro Ribeiro Barrile[3], Raquel Martins Porto[4], James Alves de Souza[5*]**

[1,2,3,5]Universidade Federal de São Carlos, UFSCar, São Paulo, Sorocaba, Brasil. mmuniz@estudante.ufscar.br, emanuele@estudante.ufscar.br, joaobarrile@estudante.ufscar.br e jasouza@ufscar.br. [4]Escola Estadual Jardim Daniel David Haddad, São Paulo, Salto de Pirapora, Brasil. raquelmartinsporto@prof.educacao.sp.gov.br.

**Resumo:** A integração da educação ambiental no currículo escolar pode ser inspiradora para os estudantes assimilarem conceitos e métodos científicos apresentados nas disciplinas de Ciências de maneira contextualizada e interdisciplinar. Para efetivação desta integração em sala de aula é importante que aja uma articulação entre os cursos de formação de professores e atividades práticas sobre o tema que forneçam condições para que os licenciandos possam testar hipóteses, metodologias pedagógicas e avaliar a viabilidade das mesmas diante das limitações reais do ambiente escolar durante a sua formação. O Programa Institucional de Bolsas de Iniciação à Docência (PIBID) é uma política governamental brasileira desenvolvida para melhorar a formação de discentes de cursos de licenciatura e fornece um cenário ideal para isso. Neste trabalho os estudantes do PIBID desenvolveram a proposta através do processo de compostagem comum. Este foi executado na escola considerando a separação seletiva do lixo, a construção das composteiras, introdução e discussão de conceitos científicos e análise dos resultados, com o monitoramento do processo. Além da experiência benéfica para a formação dos licenciandos, mostramos que a proposta é promissora para o desenvolvimento de habilidades como o letramento científico, o protagonismo dos estudantes e o estabelecimento de um programa de compostagem escolar.

**Palavras-chave:** compostagem, interdisciplinaridade, colaboratividade, ensino de ciências.

**Title:** Encouraging the teaching of science through composting

**Abstract:** The integration of environmental education into the school curriculum can be inspiring for students to assimilate scientific concepts and methods presented in the Science subjects in a contextualized and interdisciplinary way. To carry out such integration in the classroom it is important that a link between teacher training courses and practical activities on the subject during the training of the student teachers exist, to provide them conditions for testing hypotheses, pedagogical methodologies and evaluate their feasibility, considering real limitations of the school environment. The Institutional Program of Scholarships for Teaching Initiation (PIBID - Programa Institucional de Bolsas de Iniciação à Docência) is a Brazilian government policy developed to improve the training of student teachers which provides an ideal scenario for this. In this work, students from PIBID carried out this proposal through the windrow composting process. This was developed at the school considering the selective separation of waste, the construction of compost bins, the introduction and discussion of scientific concepts and experimentation, by monitoring the process. In addition to the beneficial experience for the training of student teachers, we show that the proposal is promising for developing student's skills such as scientific literacy and protagonism, and also the establishment of a school composting program.

**Introdução**

Estudos têm demonstrado que a incorporação da educação ambiental no currículo escolar tem favorecido a melhoria do comportamento e o desempenho dos estudantes (Bodzin, Klein e Weaver, 2010; Duhll e Verma, 2017). Contudo, no estado de São Paulo, Brasil, este assunto é usualmente abordado teoricamente no currículo escolar das áreas de Ciências da Natureza através de temas relacionados a ecossistemas, biodiversidade e questões ambientais, como energia, mudanças climáticas, poluição e recursos naturais (São Paulo, 2023). Para que os estudantes tenham um contato mais efetivo, do ponto de vista científico, com assuntos referentes ao meio ambiente, é importante que os cursos de formação de professores promovam atividades práticas sobre o tema, fornecendo condições para que os licenciandos possam testar hipóteses, metodologias pedagógicas e avaliar a viabilidade das mesmas nas condições e limitações reais do ambiente escolar durante a sua formação.

O Programa Institucional de Bolsas de Iniciação à Docência (PIBID) fornece um cenário ideal para fortalecer essa prática e auxiliar os futuros professores a terem mais confiança e disposição para incorporar práticas pedagógicas e atividades de aprendizagem sobre educação ambiental e ciência em sala de aula. O PIBID é uma política governamental brasileira com a finalidade de melhorar a formação de discentes de cursos de licenciatura de instituições de educação superior federais, estaduais, municipais e comunitárias sem fins lucrativos, com projetos de iniciação à docência destinados à valorização do magistério e ao aperfeiçoamento da formação dos educadores para a elevação da qualidade da educação básica (Brasil, 2024). Este programa é administrado pela Coordenação de Aperfeiçoamento de Pessoal de Nível Superior (CAPES) e integra a Política Nacional de Formação de Professores do Ministério da Educação.

Neste trabalho apresentamos uma proposta para promover tal contato, dos licenciandos com assuntos sobre o meio ambiente, considerando o tema compostagem vinculado ao currículo de Ciências da Natureza. Nosso objetivo é conscientizar toda a comunidade escolar para questões ambientais, como reciclagem, sustentabilidade e melhor gerenciamento do lixo que produzimos. A proposta foi desenvolvida em uma escola pública do interior de São Paulo por estudantes do PIBID em um subprojeto de Ciências da Natureza com um núcleo multidisciplinar, envolvendo licenciandos dos cursos de Biologia, Química e Física.

A prática da compostagem evoluiu durante o século XX de maneira sistematizada através de estudos científicos, surgindo no século XXI como uma tecnologia que pode ser essencial para a gestão de resíduos orgânicos e o desenvolvimento sustentável (Epstein, 1997). Diante dessa potencialidade é imprescindível que a mesma esteja mais presente no currículo escolar como uma alternativa de ensino que seja capaz de fortalecer a ponte entre os conhecimentos da educação ambiental com aqueles provenientes das áreas de Ciências da Natureza. Adicionalmente, a prática da compostagem na escola pode auxiliar na sensibilização e na compreensão dos fenômenos e processos que influenciam e alteram o meio ambiente e no estabelecimento de projetos escolares colaborativos, para que os estudantes possam entender questões ambientais mais complexas, avaliar leis e planos ambientais propostos e a compreender como decisões pessoais podem afetar significativamente o ambiente em uma escala local e global.

Nas próximas seções apresentamos sugestões de como o tema compostagem pode enriquecer o currículo escolar das áreas de Ciências da Natureza com atividades teóricas e práticas, visando trabalhar as três metodologias da ciência: a observação, a experimentação e as análises matemáticas. Toda a abordagem foi desenvolvida para estimular o ensino de Ciências de maneira interdisciplinar, fornecendo contextos adequados para o professor trabalhar conceitos inerentes das áreas de Biologia, Física e Química. Apresentamos também algumas ideias para implementar um programa de compostagem na escola de maneira simples e sustentável, considerando a experiência que tivemos no PIBID durante a aplicação de nossa proposta.

Esperamos que a comunidade escolar reflita sobre suas atividades e suas atitudes com relação ao meio ambiente, administrando e transformando o lixo que produz de maneira responsável e sustentável através de investigações científicas, para que o estudante possa trabalhar o letramento científico, o seu protagonismo no processo ensino-aprendizagem e na participação, proposta e desenvolvimento de projetos colaborativos escolares.

**Incorporando a compostagem ao currículo escolar de Ciências da Natureza**

Todos os conceitos e métodos apresentados nesta seção são referentes principalmente ao processo de compostagem em leiras, usualmente conhecido como compostagem comum. Neste, a decomposição da matéria orgânica é realizada apenas por microrganismos decompositores, como bactérias e fungos, sem a intervenção de vermes, como minhocas e piolhos-de-cobra (Inácio e Miller, 2009; Oshins e Michel, 2021).

A compostagem é um processo essencialmente aeróbico e termofílico de biodecomposição da matéria orgânica. Isso significa que os microrganismos responsáveis pelo processo utilizam o oxigênio ($O_2$) do ar enquanto consomem a matéria orgânica presente nos restos de alimentos utilizados como matéria-prima. A atividade metabólica dos microrganismos gera energia, fazendo com que a temperatura do composto atinja valores entre 50 e 70 ºC. Durante este processo há também a geração de dióxido de carbono ($CO_2$) e vapor d'água, que são liberados no ar. As perdas de água podem equivaler a mais da metade do peso da matéria-prima original utilizada. Dessa forma, o processo de compostagem reduz o volume e a massa do material inicial, enquanto o transforma no composto que será utilizado como adubo para o solo (Inácio e Miller, 2009; Oshins e Michel, 2021).

A matéria-prima fornece nutrientes e energia para os microrganismos decompositores, que estão presentes naturalmente na própria matéria-prima e em sua vizinhança. Estes são conhecidos como microrganismos nativos. A compostagem pode ser mais rápida quando condições que favorecem o crescimento desses microrganismos são estabelecidas, como a utilização de materiais orgânicos dispostos em pedaços com tamanhos adequados e devidamente misturados, relação carbono e nitrogênio (C:N) balanceada, pH, umidade, temperatura e oxigênio a níveis adequados (Oshins e Michel, 2021).

Note a infinidade de conceitos científicos que podem ser abordados apenas com a definição de compostagem. Para entender melhor como alguns destes conceitos podem ser trabalhados em sala de aula, considerando as três metodologias científicas, vamos ampliar algumas definições e processos apresentados nos dois parágrafos anteriores.

A discussão científica pode ser conduzida com perguntas norteadoras como: o que é biodegradação e biodecomposição? Essa é uma questão inicial muito importante para fornecer aos estudantes a escala temporal da natureza para a reciclagem natural do lixo que produzimos e o tipo de material que podemos utilizar em uma composteira.

A biodegradação é um processo de biodecomposição que envolve transformações químicas conduzidas pela ação de microrganismos. Contudo, para que um material seja considerado biodegradável é necessário que o mesmo se decomponha em semanas ou meses. Materiais sintéticos como latas de alumínio, vidro, borracha, lâmpadas fluorescentes, plásticos derivados do petróleo, lenços umedecidos, etc., são exemplos de materiais não biodegradáveis, pois podem levar décadas para se decomporem naturalmente. Isso justifica a necessidade da separação adequada do lixo para a compostagem. Outras questões ambientais ainda podem ser levantadas nesse tema de maneira complementar, como a possibilidade de compostagem de plásticos biodegradáveis, alternativas possíveis para minimizar a produção de lixo, o que é ser ecologicamente correto, dentre outras (São Paulo, 2023).

Na sequência o professor pode abordar tanto os processos envolvidos na biodecomposição da matéria-prima quanto a classificação e a caracterização dos microrganismos responsáveis pelos mesmos, com outras perguntas como: quais microrganismos estão presentes no processo de compostagem? O que são processos aeróbico e termofílico?

O processo de compostagem é iniciado imediatamente após a matéria orgânica ser colocada na composteira em condições específicas. Os microrganismos decompositores são compostos por bactérias e fungos já existentes nas superfícies de cada partícula do material e há a disponibilidade de muito oxigênio no ar. Como o processo é aeróbico as bactérias e os fungos utilizam oxigênio enquanto consomem açúcares, amidos e proteínas disponíveis como alimento para obtenção de energia e dos compostos necessários para o crescimento e o metabolismo dos mesmos (Epstein, 1997; Oshins e Michel, 2021). O professor pode discutir sobre a morfofisiologia e a reprodução desses microrganismos e como se dá o metabolismo dos mesmos através de processos bioquímicos, além de contextualizar melhor o que é matéria orgânica e inorgânica através de compostos químicos (Carneiro e Junqueira, 2012; Russel, 1994).

A temperatura é um parâmetro físico essencial no processo de compostagem. A atividade microbiana faz com que a temperatura da massa bruta no interior da composteira aumente. A fonte original dessa energia é a transformação da energia solar em energia química durante a fotossíntese realizada nas plantas. A temperatura continua a aumentar até que as substâncias orgânicas mais facilmente degradáveis sejam consumidas e as populações microbianas de espécies mais adaptadas para metabolizar os compostos orgânicos mais difíceis aumentam. As diferentes temperaturas fazem com que o ambiente fique impróprio para alguns microrganismos e mais favorável para outros. Aqueles que são inibidos ou morrem tornam-se alimentos para aqueles que prosperam. O processo global entre os diferentes conjuntos de microrganismos é complexo e dinâmico e isso afeta a taxa de decomposição. O aumento de temperatura permite classificar os microrganismos associados com a compostagem em mesofílicos e termofílicos (Epstein, 1997).

Os microrganismos ativos em todo o processo de compostagem são descritos principalmente por bactérias e fungos. Os agentes mesofílicos predominam no processo a temperaturas em torno de 40 ºC, enquanto que os termofílicos são predominantes a temperaturas maiores, de 45 a 65 ºC. A fase mesofílica pode levar de 15 horas a 3 dias, dependendo do tipo de material orgânico utilizado na composteira e outras condições já mencionadas, como o oxigênio disponível, relação carbono e nitrogênio e umidade. Já a fase termofílica pode durar várias semanas e é caracterizada pela intensa decomposição da matéria contida na composteira. À medida que a atividade dos microrganismos diminui, a temperatura do composto diminui gradualmente para 40 ºC, reativando a fase mesofílica, caracterizada pela degradação de substâncias orgânicas mais resistentes e perda de umidade. A temperatura continua diminuindo até atingir a fase de maturação, em que há grande formação de substâncias húmicas e baixa atividade biológica. Ao final desta fase o composto resultante é estável, estando pronto para ser utilizado como fertilizante para o solo (Inácio e Miller, 2009; Oshins e Michel, 2021).

Um fato muito interessante no processo de compostagem é que este permite a eliminação de microrganismos patogênicos, que são aqueles que podem causar doenças nos seres humanos, como a salmonela, e também inibe a proliferação de larvas de moscas. Este último ocorre na faixa de temperatura de 55 a 70 ºC (Epstein, 1997).

O monitoramento da temperatura da composteira permitirá aos estudantes estabelecer padrões característicos que fornecem informações sobre as mudanças na taxa e no tipo de decomposição que está ocorrendo durante o processo de compostagem. Dependendo da estabilidade e qualidade do composto resultante eles podem estabelecer relações do mesmo com deficiências de nutrientes para o melhor desenvolvimento dos microrganismos decompositores, devido à má aeração e baixa umidade da matéria-prima utilizada, por exemplo. Estas podem ser corrigidas para otimizar o processo com o reabastecimento de oxigênio, através do revolvimento da matéria orgânica e adição de água ou matéria-prima úmida, respectivamente. É importante ressaltar que essas correlações só podem ser realizadas através de um processo de investigação científica criterioso e que pode ser conduzido totalmente pelos estudantes.

O professor pode ainda aproveitar a oportunidade para fazer analogias para fornecer um melhor entendimento aos estudantes sobre temas amplamente discutidos nas mídias e de extrema importância para a nossa sobrevivência, como o aquecimento global. A conclusão sobre a existência do aquecimento do planeta só é possível porque os cientistas monitoram a temperatura do mesmo há décadas. Assim como acontece com alguns microrganismos no processo de compostagem, se a temperatura do planeta continuar aumentando, nós não teremos condições de sobreviver, devido aos problemas de saúde que poderão ser desencadeados pelas altas temperaturas, como desidratação extrema, problemas cardiovasculares, respiratórios, entre outros, ou pela escassez de alimentos de origem vegetal e animal que consumimos, que também não poderão prosperar em condições adversas de temperatura.

O processo de aquecimento da composteira também oferece ao professor condições ideais para introduzir, discutir e diferenciar conceitos de Física, como temperatura, calor e energia. Todos os objetos nos estados sólido, líquido e gasoso, são compostos por uma quantidade muito grande de átomos e moléculas que se movem continuamente de maneira desordenada. Esse movimento é conhecido como movimento térmico ou, mais comumente, como agitação térmica. Quanto mais quente for um corpo, maior será a amplitude desse movimento ou dessa agitação. A temperatura mede a intensidade da agitação térmica dos átomos e moléculas que constituem um corpo. Para toda matéria massiva existente no universo, composta por um grande número de partículas constituintes, sejam estas átomos ou moléculas, é possível estabelecer um valor de temperatura para a mesma (Bazarov, 1964).

Diferentemente da temperatura, o calor está relacionado a um processo e não pode ser aferido com nenhum aparelho de medida. Isso significa que não é possível medir o calor de um corpo para nos referirmos ao estado térmico do mesmo, se este está quente ou frio. Para isso, nós medimos a temperatura. O calor é um processo pelo qual energia é transferida de uma parte para outra de um corpo, ou entre corpos diferentes, sempre que há diferença de temperatura entre os mesmos. O fluxo de energia nesse processo é estabelecido naturalmente do corpo de maior temperatura para o corpo de menor temperatura. Isso significa que se quisermos aquecer um objeto, precisamos colocar o mesmo em contato térmico com outro a temperatura maior, ou seja, o objeto mais quente irá esquentar o objeto mais frio. Isso ocorre porque o primeiro transfere energia através de calor para o segundo. À medida que o tempo passa e energia é transferida entre os dois corpos através de calor, estes atingirão o equilíbrio térmico, o qual ocorre quando os objetos estão à mesma temperatura. Este estado de equilíbrio pode ser monitorado com um termômetro e quando o mesmo é atingido, o processo calor é interrompido, porque neste estado não há fluxo líquido de energia entre um sistema e outro. Portanto, o calor é um processo que só pode ser estabelecido quando há diferença de temperatura entre dois sistemas. O uso da palavra calor no nosso cotidiano é inadequado do ponto de vista científico, pois o mesmo se refere usualmente às nossas sensações térmicas, que são imprecisas, ou a uma substância que um corpo transmite para outro e não, necessariamente, a um processo de transmissão de energia (Silva, Laburu e Nardi, 2008).

Observe que, sempre que calor é mencionado nos referimos a um processo de transmissão de energia. Mas o que é energia?

A energia é a medida da capacidade de a matéria interagir com a matéria, tendo como resultado mudanças nas propriedades físicas dos corpos envolvidos na interação (Hecht, 2019). Isso significa que qualquer parâmetro físico de um corpo, como a temperatura da composteira, por exemplo, pode ser um indicador de que o mesmo está interagindo com sua vizinhança ou que a interação está ocorrendo entre as suas partes. Ou seja, sabemos que as interações estão ocorrendo quando verificamos, através de medidas, que estes parâmetros estão variando com o tempo, pois isso é uma consequência da transmissão de energia entre as partes interagentes.

No caso da compostagem é possível verificar o aquecimento do sistema, mesmo que este não esteja interagindo com um agente externo para produzir esse efeito. Ninguém está aquecendo o composto. Esse aquecimento é a manifestação da energia produzida pelos microrganismos decompositores devido à interação dos mesmos com a matéria orgânica depositada na composteira e pode ser monitorado pelos estudantes através da medida da temperatura com um termômetro em graus Celsius (ºC) ou outra escala de medida, como a Fahrenheit (ºF) ou a Kelvin (K).

Se não for possível medir qualquer alteração nos parâmetros físicos de um sistema no decorrer do tempo, como a temperatura, o volume, a pressão, a densidade, etc., significa que o sistema está isolado e a interação do mesmo com a sua vizinhança, ou entre as suas partes, não está acontecendo de maneira perceptível ou significativa.

As discussões científicas podem continuar a ser conduzidas tendo como tema as condições para que o processo de compostagem seja bem sucedido, considerando a aeração, a umidade, o tamanho dos pedaços dos materiais orgânicos disponíveis, a relação carbono/nitrogênio (C:N), o pH do composto, etc.

A aeração garante que o ar circule por toda a matéria-prima depositada na composteira para abastecer a necessidade de oxigênio dos microrganismos decompositores, uma vez que a compostagem é um processo essencialmente aeróbico. Na compostagem comum, esse processo pode ser estabelecido naturalmente através de correntes de convecção e difusão molecular, ou manualmente, pela mistura periódica da matéria-prima no interior da composteira.

As correntes de convecção é outro conceito inerente da Física que o professor pode trabalhar de maneira contextualizada através da compostagem. Este ocorre devido às diferenças de temperatura em um fluido e provoca deslocamento de massa. O fluido presente na composteira é o ar, que é um gás. Este diminui de densidade à medida que a sua temperatura aumenta. O campo gravitacional, sob o qual todos os corpos no planeta estão sujeitos, faz com que a porção de ar mais densa e de menor temperatura seja movida para baixo, ocupando o lugar da porção de ar menos densa e de maior temperatura que está em contato com o composto. Os fluxos ascendente e descendente de massa de ar é o que formam as correntes de convecção. É essencial que essa corrente de ar, ou aeração, seja estabelecida, pois o gás em contato com o composto torna-se rico em $CO_2$ e pobre em $O_2$ devido à atividade microbiana, que é caracterizada pelo consumo de oxigênio e emissão de gás carbônico e vapor d'água. Contudo, esse processo natural pode ser demorado e ineficiente, dependendo da forma como a composteira é construída e o volume de matéria orgânica que é depositada na mesma.

Uma vez que o composto é constituído por diferentes materiais, há a formação de espaços livres entre as partículas do mesmo, conhecidos como poros. Dependendo da profundidade da composteira, as trocas gasosas das camadas inferiores serão feitas preferencialmente por difusão molecular através desses poros, que é um processo bem mais lento que a convecção, pois este é conduzido pelo movimento térmico das moléculas. Devido a isso, é preferível que a aeração do composto seja feita manualmente, pela mistura periódica do mesmo. Para facilitar este processo, sugerimos que as composteiras escolares sejam construídas com profundidades menores, em torno de 40 a 50 cm. Forneceremos mais informações sobre a construção da composteira na próxima seção.

A falta de oxigênio no processo de compostagem irá favorecer condições anaeróbicas. Consequentemente, compostos putrescíveis, sujeitos a fermentação, serão formados, de maneira que odores fortes e desagradáveis serão percebidos (Epstein, 1997). A partir desse contexto, o professor pode diferenciar microrganismos e processos aeróbicos e anaeróbicos, falar sobre a eficiência dos mesmos na decomposição, explicar o motivo de lixões e fossas sépticas emitir odores desagradáveis, entre outras. Além de temas sobre ecologia, o professor pode explorar questões sobre sustentabilidade e educação ambiental de maneira geral, como a importância de separar o lixo orgânico para o descarte e manuseio correto do mesmo através da compostagem. O lixo orgânico produzido por nós representa em torno de 45 a 60% em peso de todo o lixo que vai para aterros sanitários, lixões a céu aberto ou depósitos irregulares, como beiras de rios e áreas alagáveis. Nestes locais os processos anaeróbicos são predominantes favorecendo a produção de uma variedade de compostos intermediários voláteis, como o gás metano, que contribui para o efeito estufa, o gás sulfídrico, que é um gás poluente que participa da formação de chuvas ácidas, ácidos orgânicos, dentre outras substâncias. Além de produzir fortes odores, ser uma fonte de poluição de recursos hídricos, fornecer um meio de proliferação de insetos vetores de doenças, alguns destes compostos são fitotóxicos, ou seja, são tóxicos para as plantas (Inácio e Miller, 2009; Oshins e Michel, 2021).

A umidade é um dos fatores mais importantes no processo de decomposição de materiais orgânicos, pois a água é essencial para a atividade biológica. A umidade que envolve as partículas do composto fornece o meio necessário para os processos metabólicos dos microrganismos, permitindo a dissolução e transporte de nutrientes, reações químicas e a locomoção dos mesmos. Portanto, se a matéria-prima no interior da composteira não tiver umidade suficiente, a atividade biológica será retardada significativamente. Por outro lado, se a quantidade de humidade for muito alta, os poros poderão ficar completamente preenchidos com água comprometendo a aeração do composto e o transporte de oxigênio. A quantidade de umidade recomendada no processo de compostagem é de 40 a 65%, mas esta depende do tipo de matéria-prima que está sendo decomposta e do estágio de decomposição (Inácio e Miller, 2009; Oshins e Michel, 2021).

A partir dessa discussão o professor pode chamar a atenção dos estudantes para o tamanho ideal dos pedaços dos materiais orgânicos disponíveis. Água e oxigênio devem estar presentes em quantidades abundantes no processo de compostagem para favorecer a atividade dos microrganismos aeróbicos. Para que isso seja estabelecido as partículas do composto não podem ser muito grandes, pois estes levam muito tempo para se decomporem, mas também não podem ser muito pequenas, para evitar a formação de poros pequenos que são facilmente preenchidos com água, dificultando a aeração e o transporte de oxigênio.

A atividade dos microrganismos aeróbicos se dá majoritariamente a partir da superfície das partículas do composto. Quando cortamos a matéria orgânica, como uma fatia grande de casca de melancia, por exemplo, em pedaços menores, nós estamos disponibilizando uma superfície maior para a ação dos microrganismos, o que irá acelerar o processo de compostagem. À medida que a atividade microbiana se intensifica, a umidade do composto aumenta formando um filme de água cada vez mais espesso em torno de cada partícula do composto, o que dificulta a difusão do oxigênio no filme de água até chegar no substrato sólido, onde estão os microrganismos decompositores. A falta de oxigênio favorece a atividade de microrganismos anaeróbicos e o surgimento de odores e substâncias tóxicas indesejadas.

Para evitar a formação de filmes cada vez mais espessos e o acúmulo de água, a composteira deve ser construída de uma maneira que a água em excesso possa escoar para fora do composto. Isso é facilitado consideravelmente quando o composto é misturado periodicamente. Portanto, parte do trabalho dos estudantes pode ser investigar cientificamente, através dos métodos da observação e da experimentação, a quantidade ideal de umidade no interior da composteira para o tipo de matéria orgânica utilizada e para as condições de compostagem estabelecidas. Essa análise pode ser simplificada se eles souberem de antemão que muita umidade irá interferir na aeração do composto, conduzindo a condições anaeróbicas, e pouca umidade irá comprometer consideravelmente a atividade microbiana responsável pela compostagem.

Para atingir a umidade mais adequada, os estudantes podem realizar testes da proporção da mistura de materiais mais secos, como folhas secas, restos de poda e de jardinagem, com materiais mais úmidos, como cascas e sobras de frutas. É interessante que o professor incentive os estudantes a escreverem relatórios ou diários de bordo durante o monitoramento do processo para registrar variações de parâmetros, formular perguntas, apresentar hipóteses e conclusões prévias, com o objetivo de aperfeiçoar a escrita acadêmica e desenvolver o senso crítico dos estudantes.

Além da umidade e da oxigenação, que fornecem o meio de cultura para os microrganismos decompositores, é importante avaliar o tipo de nutrientes que os mesmos necessitam. Dentre os principais estão o carbono (C), o nitrogênio (N), o fósforo (P), o potássio (K) e o enxofre (S). Apesar de C e N serem os dois nutrientes mais importantes nesse processo, é indispensável que a presença de P e K seja considerada, pois estes, junto com N, são nutrientes essenciais para as plantas, de maneira que a concentração dos mesmos agrega valor ao composto final como fertilizante.

Os resíduos vegetais e alimentares, que são os que utilizamos na composteira escolar, contém grandes quantidades desses nutrientes para alimentar os microrganismos. Uma quantidade balanceada de carbono e nitrogênio, usualmente referida como razão C:N, garantem que os outros nutrientes requeridos estejam presentes no composto (Epstein, 1997; Oshins e Michel, 2021). Neste contexto, o professor pode explorar a bioquímica, através do metabolismo dos microrganismos decompositores, a química orgânica, com o estudo do elemento carbono, compostos orgânicos nitrogenados e sulfurados, a biogeoquímica, com o ciclo do carbono na natureza, além de outros conceitos como a densidade volumétrica e o potencial hidrogeniônico (pH) do composto.

A fase final do processo de compostagem é chamada de maturação. Esta etapa ocorre a temperaturas mesofílicas mais baixas e é caracterizada pelo decréscimo monotônico da temperatura em função do tempo até que a temperatura ambiente (~ 25 ºC) seja atingida (Oshins e Michel, 2021). Se a temperatura da composteira for monitorada periodicamente pelos estudantes, eles poderão determinar o início desta fase, pois como a atividade microbiana diminui consideravelmente, não há o reaquecimento do composto quando este é revirado manualmente. Isso significa que a temperatura irá apenas decrescer em função do tempo, ou seja, esta apresentará um decréscimo monotônico.

A fase de maturação foi a mais demorada no processo de compostagem escolar e pode durar em torno de 2 meses, dependendo do tipo e do preparo da matéria-prima utilizada e do tamanho da composteira. O maior tempo é justificado porque a maturação promove a decomposição aeróbica de compostos mais resistentes, como ácidos orgânicos, partículas maiores e aglomerados de material que permaneceram após o período mais ativo da compostagem. Alguns fungos e bactérias, como os actinomicetos, são ativos apenas em temperaturas mesofílicas de maturação para a decomposição de compostos celulósicos.

Mas como os estudantes poderão avaliar quanto tempo a etapa de maturação deve durar e quando o produto final poderá ser utilizado como fertilizante?

Essa avaliação é muito importante porque o composto imaturo continua consumindo oxigênio e isso reduzirá a disponibilidade desse gás para as raízes das plantas, podendo prejudicá-las. Este também pode conter altos níveis de ácidos orgânicos e amônio, além de outras substâncias fitotóxicas (Oshins e Michel, 2021). Os principais indicadores que observamos na compostagem comum para saber se o composto está pronto para ser testado e utilizado como fertilizante foram o odor e a aparência do mesmo. Se o composto apresentar qualquer tipo de odor característico ou se restos de alimentos ainda forem perceptíveis é necessário um período maior de maturação. Se o mesmo apresentar uma aparência uniforme com coloração escura e sem cheiro, significa que o produto está estável, com baixa atividade microbiana e uma boa concentração de substâncias húmicas. Contudo, a avaliação mais precisa para saber se o composto produzido atingiu a maturidade necessária para o tipo de plantas que a comunidade escolar deseja cultivar é através de bioensaios.

Além do adubo sólido nós obtemos também o chorume, como resultado da perda de água do composto devido à atividade microbiana. Este produto é líquido e composto por uma quantidade muito grande de microrganismos benéficos, como bactérias e fungos, além de ser rico em nutrientes solúveis, podendo ser utilizado para beneficiar as plantas diretamente no solo ou como adubo foliar, ou seja, pulverizado sobre as folhas.

Quando aplicado na superfície do solo tanto o chorume quanto o composto sólido fornecem nutrientes às raízes das plantas para melhorar o seu crescimento, com efeitos positivos na estruturação do solo, na retenção de água, na profundidade das raízes, na melhoria da ciclagem e retenção de nutrientes e auxilia a desenvolver uma barreira biológica em torno das raízes agindo na supressão de doenças (Ingham, 2005). Quando o chorume é aplicado como adubo foliar, os nutrientes são absorvidos pela parte aérea das plantas, principalmente as folhas, e são mais rapidamente assimilados pelas mesmas do que quando aplicado no solo.

Se a escola possuir um peagâmetro, os estudantes podem medir o pH do chorume obtido. De maneira geral, uma solução nutritiva possui pH entre 5 e 6 (Matioli, 2019). O professor pode aproveitar o contexto de adubação e apresentar a morfologia das plantas para explicar, por exemplo, a penetração dos nutrientes nas folhas através dos estômatos e como se dá a distribuição dos mesmos das folhas para outras partes das plantas. É importante salientar que a adubagem foliar não substitui a adubagem do solo, ambas se complementam. Dessa forma, o processo de compostagem escolar pode fornecer condições ideais para a produção de adubo orgânico para manter a escola bonita e sustentável com a plantação de hortaliças, jardins e outras plantas.

Do ponto de vista de análises matemáticas, o professor de qualquer área da ciência também pode se beneficiar consideravelmente com as atividades de compostagem. As composteiras podem ser feitas utilizando baldes ou caixas plásticas. Estes possuem uma forma geométrica espacial com volume bem definido. Neste contexto, o professor pode trabalhar tanto geometria plana quanto geometria espacial com os estudantes.

Se considerarmos um balde cilíndrico de altura h, por exemplo, o seu volume V pode ser calculado através da expressão $V = Ah$, sendo $A = \pi R^2$ a área de sua base, que é um círculo de raio R. Os estudantes podem também fazer marcações no balde com diferentes alturas, $h_1$, $h_2$, $h_3$, etc., para calcular o volume respectivo de matéria-prima que está sendo depositada diariamente na composteira e estimar a quantidade de lixo orgânico que é produzido na escola ou em suas casas neste período.

Adicionalmente, eles podem comparar o volume ou a massa inicial de matéria-prima utilizada na composteira com o volume ou a massa final de composto produzido para estimar qual foi a perda em volume ou em massa do material original com o processo de compostagem. Conhecendo-se a massa m do composto eles também podem calcular sua densidade pela equação $d = m/V$. Estas poucas equações já permitem introduzir unidades de medida, seus múltiplos e conversões, como centímetros (cm) ou metros (m) para medidas de comprimento, largura e altura; centímetros cúbicos ($cm^3$), metros cúbicos ($m^3$) ou litros (l) para medidas de volume; quilogramas (kg) e gramas (g) para medidas de massa e as combinações destas para outras medidas, como a densidade dada em $kg/m^3$ ou $g/cm^3$.

O professor pode ainda introduzir gráficos para analisar como a temperatura do composto varia com o tempo, calcular razões para estabelecer relações entre grandezas, como a própria densidade, a velocidade ou a taxa de decomposição da matéria-prima e também calcular proporções para comparar as densidades final e inicial do composto, por exemplo.

Para estimular o raciocínio crítico dos estudantes, situações-problema que envolvem o conhecimento matemático para resolvê-las podem ser criadas, como por exemplo: conhecendo-se a massa inicial e final do composto em um processo de compostagem, qual deve ser a razão entre os volumes final e inicial, sabendo-se que a densidade do produto final é o dobro da inicial?

Outra forma interessante de trabalhar as análises matemáticas é através de equações químicas, como o processo básico de respiração dos microrganismos decompositores. Se considerarmos a glicose ($C_6H_{12}O_6$) como o nutriente consumido durante a atividade microbiana, o processo se dá com o consumo de oxigênio ($O_2$), a emissão de gás carbônico ($CO_2$) e vapor d'água ($H_2O$) e a geração de energia. A equação química correspondente é escrita como:

$$C_6H_{12}O_6 + 6\ O_2 \rightarrow 6\ CO_2 + H_2O + \text{Energia}, \qquad (1)$$

em que os reagentes são apresentados do lado esquerdo da seta e o resultado do processo químico é apresentado do lado direito. O processo pode ainda ser ilustrado de maneira geral e qualitativa para qualquer nutriente utilizado como reagente, como:

$$\text{Nutriente (energia química)} + O_2 \rightarrow CO_2 + H_2O + \text{Energia (aumento de temperatura)}. \qquad (2)$$

Além da conservação de matéria, observada matematicamente pela mesma quantidade de carbono, hidrogênio e oxigênio dos dois lados da seta na equação (1), as equações químicas podem ser muito úteis para auxiliar os estudantes a entenderem porque o composto aquece durante o processo de compostagem. A atividade microbiana fornece energia ao composto durante a respiração como resultado da quebra das moléculas dos nutrientes disponíveis na matéria orgânica, estabelecendo um processo de transformação de energia química em energia térmica. A manifestação dessa segunda modalidade de energia é verificada pela medida da temperatura do composto. Mas isso nem sempre é simples para os estudantes processarem.

A análise matemática das equações químicas mostra claramente que a quantidade de matéria massiva é conservada no processo, pois a quantidade de elementos químicos dos dois lados da equação (1) são os mesmos. Por outro lado, quando a equação química referente ao processo de respiração é apresentada como em (1), esta não permite que os estudantes vislumbrem o conceito de energia e suas transformações diretamente, mas pode subsidiar dúvidas como: de onde surgiu a energia adicional apresentada no lado direito da equação?

Essa energia é a medida da capacidade de os microrganismos decompositores interagirem com a matéria orgânica disponível. A energia que mantém os elementos químicos C, H e O ligados na molécula de glicose é liberada quando os microrganismos quebram as ligações. Parte dessa energia é processada pelos mesmos em seu metabolismo e parte é liberada no meio, fazendo com que haja o aquecimento do composto.

O processo de investigação científica conduzido pelas metodologias da observação, experimentação e análises matemáticas nas atividades de compostagem, além de poder auxiliar os estudantes na aprendizagem de vários temas, cobrindo uma grande quantidade de conceitos e métodos que estruturam o trabalho curricular da área de Ciências da Natureza, pode facilitar no desenvolvimento de várias habilidades. Dentre estas, podemos citar o letramento científico, o protagonismo do estudante durante a investigação e reflexão para a solução de problemas atuais e o desenvolvimento de atividades colaborativas, para que o mesmo se forme como um sujeito transformador consciente, com um olhar diferenciado para práticas sustentáveis e os cuidados que devemos ter com a natureza (São Paulo, 2023).

De acordo com a professora supervisora, que foi nossa colaboradora durante a execução do PIBID na escola parceira, as atividades de compostagem podem ser utilizadas para trabalhar conceitos científicos a partir do sexto ano do ensino fundamental nas aulas de Ciências, podendo ser estendida e adaptada para o ensino médio nas aulas de Biologia de maneira interdisciplinar, considerando tópicos de Física e Química. Segundo ela, além de ser um facilitador que permite explorar a totalidade metodológica da ciência e o desenvolvimento de habilidades dos estudantes, considerando o contexto social e as relações com outras áreas do conhecimento, nossa proposta tem um grande potencial para abordar de maneira contextualizada os Objetivos do Desenvolvimento Sustentável (ODS), estabelecidos pela Organização das Nações Unidas para proteger o meio ambiente e o clima e garantir qualidade na educação e de vida para todos (UN DESA, 2021).

**Ideias para iniciar um programa de compostagem na escola**

Para a implementação e execução das atividades de compostagem é necessário planejamento, conscientização e colaboração da comunidade escolar, pois todo o processo demanda em torno de três meses. Apesar de ser um processo de longo prazo, o mesmo pode ser executado desde a elaboração das atividades, apresentação e discussão do conteúdo científico correspondente em sala de aula, separação do lixo orgânico para aquisição de matéria-prima, preparação da composteira e obtenção do adubo orgânico em um único semestre.

Uma vez estabelecido é importante que o processo seja continuado, mesmo quando os professores e os estudantes mais engajados na proposta deixem a escola, pois além de ser uma excelente alternativa para o manejo de recursos naturais, a compostagem permite que os estudantes estejam em constante contato com assuntos referentes a sustentabilidade e os cuidados que devemos ter com o meio ambiente. É desejável também que a escola forneça condições para os estudantes levarem a proposta para suas casas através de panfletos informativos simples e manuais de montagem de uma composteira caseira.

Para sustentar o programa de compostagem sugerimos as etapas a seguir, apresentando as estratégias e as atividades desenvolvidas pelos estudantes do PIBID na escola parceira.

1) Formar um grupo de compostagem;

A formação de um grupo é importante para manter e monitorar o programa de forma efetiva e consistente, mesmo nos períodos de férias escolares. Este grupo pode ser composto por professores, gestores, coordenadores, técnicos, secretários, inspetores, merendeiros, auxiliares de limpeza, estudantes e até mesmo pelos pais dos estudantes. A professora supervisora e os estudantes do PIBID formaram o grupo de compostagem da escola parceira durante a execução do programa.

2) Escolher um coordenador para o grupo de compostagem;

O coordenador deve ser uma fonte de motivação para o bom andamento do programa, ser responsável pela organização das atividades, delegar tarefas a outros membros da equipe para que estas possam ser executadas de maneira mais organizada com os estudantes, ajudar na solução rápida de imprevistos e garantir que as metas sejam alcançadas no período previsto.

Esta função deve preferencialmente ser atribuída a um dos professores responsáveis pelas atividades em sala de aula, por este deter um conhecimento amplo sobre os conteúdos científicos que podem ser trabalhados a partir do processo de compostagem, e sobre as possibilidades oferecidas pelo currículo escolar, com relação a determinação da carga horária de cada atividade, a metodologia que deve ser empregada e o processo de avaliação. Contudo, é interessante e desejável que algumas metas sejam estabelecidas em grupo para viabilizar a participação dos estudantes de maneira colaborativa.

3) Promover os benefícios da compostagem e a importância da separação correta do lixo;

O grupo de compostagem também deve ser responsável por promover a ideia na escola para viabilizar e fortalecer o programa. Primeiramente, é importante verificar se a escola tem condições de estabelecer um processo de separação correta do lixo em diferentes recipientes e se existe coleta seletiva na região em que a mesma está localizada. A implementação da separação correta do lixo na fonte é muito importante para evitar a contaminação do material orgânico com metais pesados, pedaços de embalagens e outros materiais inertes que podem comprometer a qualidade do produto final da compostagem.

Como a escola parceira não participava de nenhum programa de coleta seletiva de lixo da região, os estudantes do PIBID conseguiram que a coleta fosse feita por uma cooperativa com agendamento prévio, sempre que uma quantidade de lixo mínima fosse acumulada. Para a separação seletiva de plásticos, papéis, metais e vidros foi instalado um kit contendo 4 lixeiras coloridas, com capacidade de 60 litros cada, em um local estratégico do pátio da escola, próximo aos banheiros e ao refeitório, onde havia intensa circulação de estudantes. A separação dos resíduos orgânicos pode ser feita diretamente na cozinha escolar pelos funcionários, mas é importante que uma lixeira para essa finalidade seja instalada no pátio para que os estudantes possam descartar seus resíduos e contribuir na separação do lixo orgânico.

Em seguida, o grupo iniciou o trabalho de conscientização da comunidade através de panfletos e palestras com o objetivo de incentivar atitudes voltadas para a sustentabilidade, a reciclagem, a importância da separação e coleta seletiva do lixo, o gerenciamento dos resíduos urbanos, os benefícios que isso pode trazer para o ambiente, os procedimentos para montar uma composteira caseira e o tipo de alimentos que podem ser utilizados para abastecer a mesma. Na figura 1 apresentamos um exemplo de panfleto distribuído na escola.

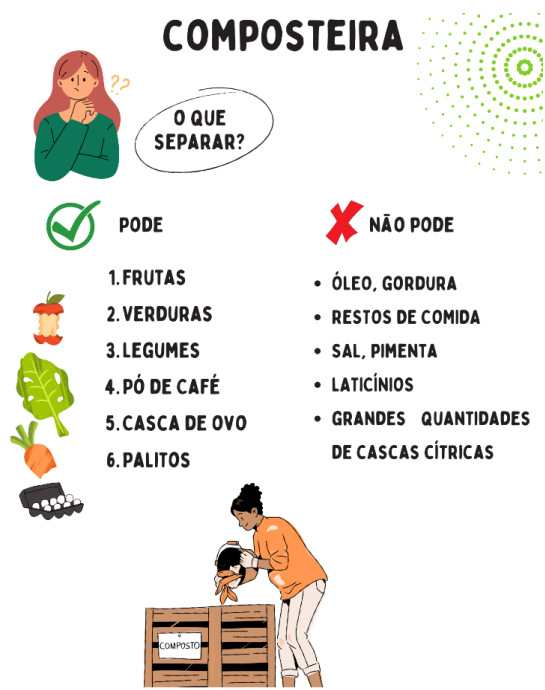

Figura 1: Panfleto sobre o tipo de alimentos que podem ser utilizados para abastecer uma composteira caseira. Elaborado pelos autores, 2023.

Para relacionar a compostagem com o tema reciclagem, que estava sendo discutido amplamente na escola, os estudantes do PIBID promoveram a mesma como uma biotecnologia ambiental responsável por reciclar os nutrientes que nos alimentam para a produção de adubo orgânico. Adicionalmente, eles divulgaram entre os estudantes que a compostagem seria integrada ao currículo escolar com atividades complementares nas aulas de Ciências e de Biologia para ser desenvolvida por eles mesmos de maneira colaborativa com a equipe escolar.

O professor pode avaliar a possibilidade dos próprios estudantes realizarem o trabalho de promoção do processo de compostagem como parte das atividades da disciplina envolvida, visando melhorar ou mesmo introduzir o letramento científico, a oralidade e a comunicação dos mesmos. Isso pode ser feito através de textos e vídeos elaborados para divulgação em redes sociais da escola com espaço para discussões, como fóruns, exposição de dúvidas, críticas e sugestões. Materiais visuais como desenhos, estampas de camisetas, adesivos, ilustrações ou tirinhas contendo dicas, incentivos e elogios sobre as atividades relacionadas podem ser confeccionados em um espaço maker, de acordo com as possibilidades e realidade de cada escola.

Para fortalecer o programa ainda mais é importante que a ideia seja também promovida fora do ambiente escolar, através de atividades de divulgação para os familiares dos estudantes utilizando boletins informativos, por exemplo, e passeios escolares que demonstrem aplicações e a importância da compostagem que está sendo desenvolvida na escola, como visitas a parques naturais, viveiros de mudas, hortas comunitárias e jardins botânicos.

4) Escolher um método de compostagem conforme a viabilidade da escola;

A compostagem no ambiente escolar pode ser feita seguindo os princípios da compostagem comum, conforme a proposta apresentada neste trabalho, ou através da vermicompostagem, com a utilização de minhocas (Ricci, 1996). As duas abordagens são igualmente enriquecedoras para a complementação do currículo escolar de ciências e podem ser desenvolvidas domesticamente, em pequenas composteiras. Contudo, no processo de vermicompostagem é necessário comprar as minhocas e fornecer condições adequadas para que as mesmas produzam o vermicomposto, conhecido como húmus de minhoca. A compostagem que realizamos não gera custo nenhum para a escola e os cuidados são reduzidos a monitorar o processo apenas, conforme discutido na seção anterior.

5) Planejar as atividades e construir as composteiras;

Uma vez que a compostagem é um processo que demanda tempo, em torno de 3 meses, é imprescindível que o professor estabeleça um cronograma de atividades para serem desenvolvidas durante todo o semestre, o que inclui a preparação do conteúdo científico relacionado, a construção das composteiras, seu abastecimento e o monitoramento do processo.

As composteiras podem ser construídas inteiramente pelos estudantes em um espaço maker. Nós utilizamos três baldes de plástico, com capacidade de 20 l cada, doados por uma casa de rações. É importante que os 3 baldes possam ser empilhados uns sobre os outros para formar uma torre com três compartimentos interconectados, dois para o depósito de resíduos orgânicos e um para o armazenamento de chorume. Optamos por utilizar baldes menores para que os estudantes pudessem ver a viabilidade de construir uma composteira caseira, que pudesse ser acondicionada facilmente em casas, apartamentos, quintais ou pequenos estabelecimentos.

Os baldes da composteira não podem ter sido utilizados previamente para armazenagem de produtos químicos, como pesticidas, tinta, materiais de construção, entre outros. Estes devem ter a finalidade única de armazenar alimentos para evitar qualquer tipo de contaminação do composto ou prejudicar o metabolismo dos microrganismos decompositores.

Antes de começar a construção das composteiras, os estudantes do PIBID elaboraram e distribuíram um zine completo e direcionado para a proposta desenvolvida na escola, contendo textos e imagens sobre o que é a compostagem, os materiais orgânicos que podem ser utilizados na composteira, os materiais e o procedimento necessário para a construção da mesma, o adubo orgânico e o chorume fertilizante provenientes do processo e como estes podem ser utilizados e armazenados. Na figura 2 à direita mostramos uma das ilustrações utilizadas no zine sobre a construção da composteira e à esquerda duas composteiras prontas utilizadas na escola.

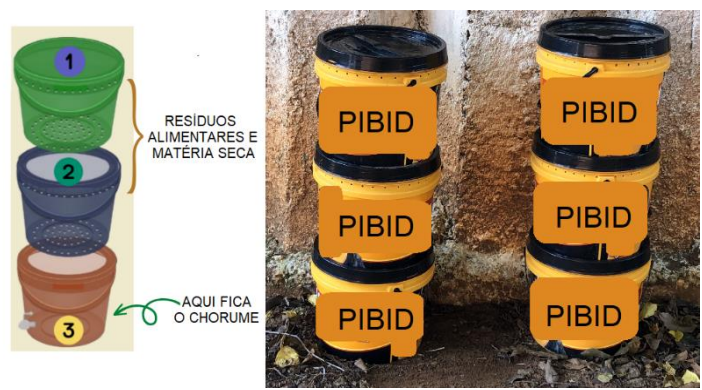

Figura 2: À esquerda mostramos uma das figuras apresentadas no zine distribuído para a comunidade escolar ilustrando como os 3 baldes devem ser dispostos uns sobre os outros e como os baldes 1 e 2 devem ser perfurados na lateral, para permitir a aeração da matéria-prima contida na composteira, e no fundo, para o escoamento do chorume produzido no processo até a sua armazenagem no balde 3. À direita mostramos uma foto das duas composteiras construídas e utilizadas nas atividades desenvolvidas na escola parceira. Elaborado pelos autores, 2023.

O modelo de composteira construído em nossa proposta é basicamente o mesmo utilizado para a vermicompostagem, exceto pela necessidade de colocar uma contenção para as minhocas na boca do balde 3. O passo a passo detalhado para a montagem da composteira, como deve ser feita a furação e o corte nas tampas dos baldes 2 e 3, seguindo a ilustração apresentada na figura 2, podem ser encontrados na referência EPAGRI (2021).

Na ilustração à esquerda da figura 2 é mostrada uma torneira instalada no balde 3 para a coleta do chorume. Como nossa proposta foi desenvolvida para evitar a geração de gastos para a escola, devido à não disponibilidade de recursos, nós não instalamos a mesma e o chorume pôde ser coletado diretamente na boca do balde sem nenhum prejuízo, bastando remover os baldes 1 e 2 para isso. A instalação da torneira é opcional.

6) Estabelecer um sistema de coleta do lixo orgânico;

Conforme já discutido, a separação adequada dos resíduos orgânicos é fundamental para o bom funcionamento do processo de compostagem e das atividades relacionadas. O grupo de compostagem da escola precisa estimar a quantidade de lixo orgânico produzida na cozinha escolar e pelos estudantes para avaliar a capacidade do sistema de compostagem construído e se há possibilidades de expandir o mesmo de acordo com a necessidade. O excesso de lixo deve ser descartado e coletado adequadamente.

A coleta deve ser feita até o preenchimento completo das composteiras. À medida que o composto orgânico estiver devidamente maturado ou ocorrer perda de volume do mesmo, devido à ação microbiana, o abastecimento das composteiras poderá ser reiniciado. É por isso que é importante que o processo de compostagem seja estabelecido de maneira continuada na escola, para que a coleta não seja feita apenas para fins didáticos, enquanto a disciplina relacionada estiver sendo ministrada. Este deve ser concebido como um processo de educação ambiental permanente.

Como nossa proposta foi desenvolvida para verificar a viabilidade do processo na escola, construímos somente duas composteiras, com capacidade total de armazenamento de apenas 40 litros. A coleta de resíduos orgânicos foi feita diariamente pelo grupo de compostagem. As composteiras foram completamente abastecidas em poucos dias. Contudo, a escola possui um espaço arborizado, protegido das chuvas, adequado e amplo para a instalação de novas composteiras. Como não há nenhum custo para a construção das mesmas, a realização da coleta de lixo e a execução da proposta, a escola tem condições de criar um sistema de compostagem com capacidade suficiente para reciclar todo o lixo orgânico produzido.

7) Executar o programa de compostagem;

Com a organização e o estabelecimento de todas as etapas anteriores o programa de compostagem pode ser executado junto aos estudantes conforme os procedimentos sugeridos na seção anterior. Na figura 3 apresentamos duas fotos tiradas durante o início das atividades do programa em sala de aula. Vários conceitos científicos foram apresentados e discutidos na prática durante a explicação de como preparar os baldes para a construção de uma composteira, a apresentação de resíduos compostáveis e como a composteira deve ser abastecida e monitorada.

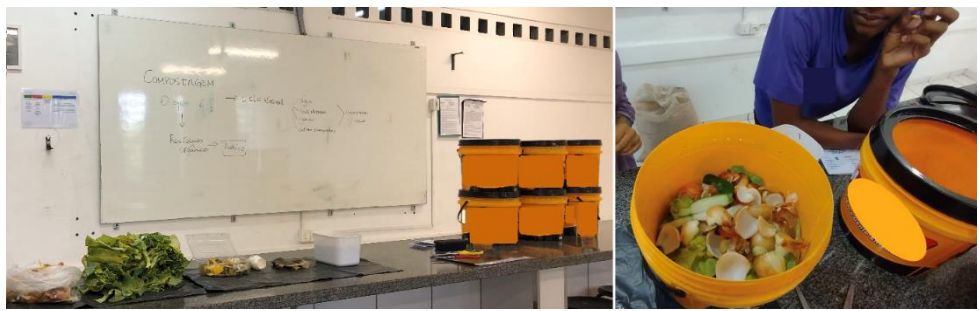

Figura 3: À esquerda é apresentada uma foto da bancada do laboratório de ensino da escola preparada para mostrar aos estudantes como construir uma composteira e com exemplos de resíduos alimentares coletados na cozinha da escola que podem ser utilizados para a compostagem comum, como restos de verduras, cascas de ovos, borra de café, cascas e restos de frutas, etc. A foto à direita mostra a atividade de abastecimento da composteira, estabelecendo inicialmente a proporção de 1:1 de matéria seca e úmida. Elaborado pelos autores, 2023.

O acompanhamento do processo de compostagem através de parâmetros como temperatura, aeração, umidade, proporção de matéria seca e úmida, entre outros, é essencial para otimização das condições necessárias para obtenção de um bom adubo orgânico, como discutido na seção anterior. Apesar das semelhanças na montagem e no abastecimento das composteiras para os processos de compostagem comum, realizado por nós, e o de vermicompostagem, o revolvimento periódico da matéria bruta no interior da composteira no primeiro caso é essencial para garantir a aeração e uma boa atividade microbiana. No processo de vermicompostagem isso é feito pelas minhocas. Devido à essa e outras diferenças é importante se atentar para a proporção de matéria seca, rica em carbono (C), e úmida, rica em nitrogênio (N), para o tamanho da composteira utilizada. No nosso caso, para cada camada de 5 cm de altura de resíduos orgânicos foi acrescentada uma camada igual de folhas secas e serragem, garantindo uma razão C:N balanceada, ausência completa de odores indesejados e um adubo orgânico de boa aparência, como apresentado na figura 4.

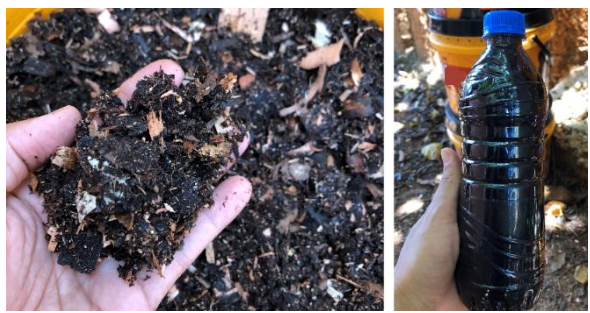

Figura 4: Fotos mostrando o adubo orgânico, sólido e líquido, obtido após o processo de compostagem comum. Elaborado pelos autores, 2023.

Pela foto apresentada na figura 4 nota-se que não é possível identificar vestígios dos resíduos orgânicos utilizados no início do processo, com exceção dos cavacos de serragem adicionados como matéria seca. Isso ocorre porque a madeira possui um polímero chamado de lignina, que é altamente resistente à degradação biológica (Oshins e Michel, 2021). O adubo sólido obtido com o nosso processo foi considerado maturado e pronto para o uso no solo após 75 dias, por possuir uma aparência homogênea, escura e um leve aroma de terra.

O chorume armazenado no balde 3 da composteira também pode ser aproveitado como biofertilizante líquido. A concentração do mesmo deve ser analisada pelos estudantes para cada tipo de aplicação desejada. Este não pode ser aplicado diariamente porque o excesso de qualquer tipo de nutriente pode prejudicar a absorção de outros nutrientes e causar deficiências nas plantas. Em um primeiro momento seguimos as recomendações para a diluição do chorume produzido na vermicompostagem, de 1 litro de chorume diluído em 10 litros de água (EPAGRI, 2021). Contudo, como a produção do chorume é diferente nos dois processos de compostagem é importante fazer uma análise através de bioensaios para otimizar a concentração e o período de aplicação do mesmo nas plantas para cada situação.

**Considerações Finais**

A proposta de desenvolver um programa de compostagem na escola pode trazer inúmeros benefícios para toda a comunidade escolar. Foi possível perceber que a gestão dos restos de comida pelos funcionários da cozinha, estudantes, professores e outros, despertou um comprometimento de todos com o ambiente escolar e a natureza. Devido à dinâmica colaborativa estabelecida nas atividades de compostagem, outros projetos começaram a ser desenvolvidos além da separação adequada do lixo produzido na escola com a instalação de lixeiras para coleta seletiva, como a construção de uma horta vertical para a plantação de temperos, uma horta horizontal para a plantação de hortaliças diversas e o revigoramento das plantas e jardins já existentes na escola com o adubo orgânico proveniente do processo.

Além de promover o desenvolvimento da comunidade, o estabelecimento de relações sociais, a colaboratividade e oportunidades para trabalhar a sustentabilidade e a educação ambiental de maneira geral, as atividades de compostagem apresentaram um potencial significativo para o professor introduzir e aprimorar diversos conceitos na disciplina de Ciências, com os estudantes do ensino fundamental, e também em áreas específicas da ciência como a Biologia, a Física e a Química, de maneira interdisciplinar, integrando os conteúdos correspondentes com a matemática, a escrita acadêmica e práticas laboratoriais.

A equipe escolar demonstrou muito interesse em manter o programa de compostagem na escola de forma continuada e dar seguimento a alguns procedimentos que não puderam ser acompanhados pelos estudantes do PIBID devido à finalização do programa, como o armazenamento do lixo seletivo até a realização da coleta pela cooperativa, a forma como a coleta seletiva será integrada à rotina escolar e se a mesma será mantida de forma consistente e estimar a quantidade de resíduos orgânicos produzidos pela comunidade escolar para realizar um planejamento adequado para a expansão do processo de compostagem.

Adicionalmente, a professora supervisora, responsável pelas disciplinas de Ciências e de Biologia, foi muito receptiva e participativa com relação às sugestões de metodologias pedagógicas, introdução de conceitos e abordagens experimentais fornecidas pela equipe do PIBID. Ela vem se dedicando, junto com outros colegas, a aprimorar a sua prática docente após a experiência vivida no programa.

**Agradecimentos**

**Referências bibliográficas**